\documentstyle[11pt,paspconf]{article}

\begin{document}

\title{Infrared Spectroscopy of the High Redshift Radio Galaxy MRC~2025-218
and a Neighboring Extremely Red Galaxy }

\author{J. E. Larkin$^1$, I. S. McLean$^1$, J. R. Graham$^2$,
E. E. Becklin$^1$, D. F.  Figer$^{1,3}$, A. M. Gilbert$^2$,
T. M. Glassman$^1$, N. A. Levenson$^{2,4}$, H. Teplitz$^{5,6}$,
M. K. Wilcox$^1$ }

\affil{$^1$Dept. of Physics and Astronomy, University of California,
Los Angeles}
\affil{$^2$Dept. of Astronomy, University of California, Berkeley}
\affil{$^3$Space Telescope Science Institute}
\affil{$^4$Dept. of Physics and Astronomy, John's Hopkins University}
\affil{$^5$LASP, Goddard Space Flight Center}
\affil{$^6$NOAO Research Associate}

\begin{abstract}
This paper presents infrared spectra taken with the newly commissioned
NIRSPEC spectrograph on the Keck Telescope of the High Redshift Radio
Galaxy MRC~2025-218 (z=2.630) and an extremely red galaxy (R-K $>$
6mag) 9'' away.  These observations represent the deepest infrared
spectra of a radio galaxy to date and have allowed for the detection
of H$\beta$, OIII (4959/5007), OI (6300), H$\alpha$, NII (6548/6583)
and SII (6716/6713).  The H$\alpha$ emission is very broad (FWHM $\sim$
6000 km/s) and strongly supports AGN unification models linking radio
galaxies and quasars.  The line ratios are most consistent with a
partially obscured nuclear region and very high excitation.  The OIII
(5007) line is extended several arcseconds and shows high velocity
clouds in the extended emission.  The nucleus also appears spectrally
double and we argue that the radio galaxy is undergoing a violent
merger process.  The red galaxy, by comparison, is very featureless
even though we have a good continuum detection in the H and K bands.
We suggest that this object is a foreground galaxy, probably at a redshift
less than 1.5.

\end{abstract}


\section{Introduction}

Deep radio surveys have proven to by one of the best methods for
finding high redshift galaxies. Most evidence points towards radio
galaxies as precursors to local giant ellipticals (e.g. Pentericci, et
al. 1999).  Many have irregular and complex morphologies suggestive of
mergers and they are often surrounded by an overdensity of compact
sources; presumably sub-galactic clumps (e.g. van Breugel et
al. 1998).  Active galaxy unification models suggest that FRII radio
galaxies are quasars with obscured broad line regions. Recent infrared
spectroscopic surveys (e.g. Evans 1998) have shown that at redshifts
less than 2.6 roughly have of the radio galaxies show evidence of an
active nucleus typically with a Seyfert 2 spectrum.

MRC 2025-218 (z=2.630) has a compact infrared and optical continuum
morphology (van Breugel et al. 1998), but extended Ly$\alpha$ emission (5'')
aligned with its radio axis (McCarthy et al. 1992). McCarthy et al. also
found three extremely red galaxies (ERO's: R-K $\>$ 6 mag) within 20''
of the radio galaxy.  This is a large overdensity of such objects and strongly
suggests and association between the ERO's and the active galaxy.

\section{Observations}

The field of MRC 2025-218 was observed on 4 Jun, 1999 (UT) with the
near infrared spectrograph NIRSPEC (McLean, et al. 1998) on the Keck
II Telescope during its commissioning.  First the field was imaged in
the K band with the slitviewing camera. Figure 1 shows the reduced
image of the field with a total integration time of 540 seconds. As
shown in the figure, the slit (42'' long and 0\farcs57 wide) was
placed on both the radio galaxy and the extremely red galaxy dubbed
ERO-A by McCarthy et al. (1992).

\begin{figure}
\plotfiddle{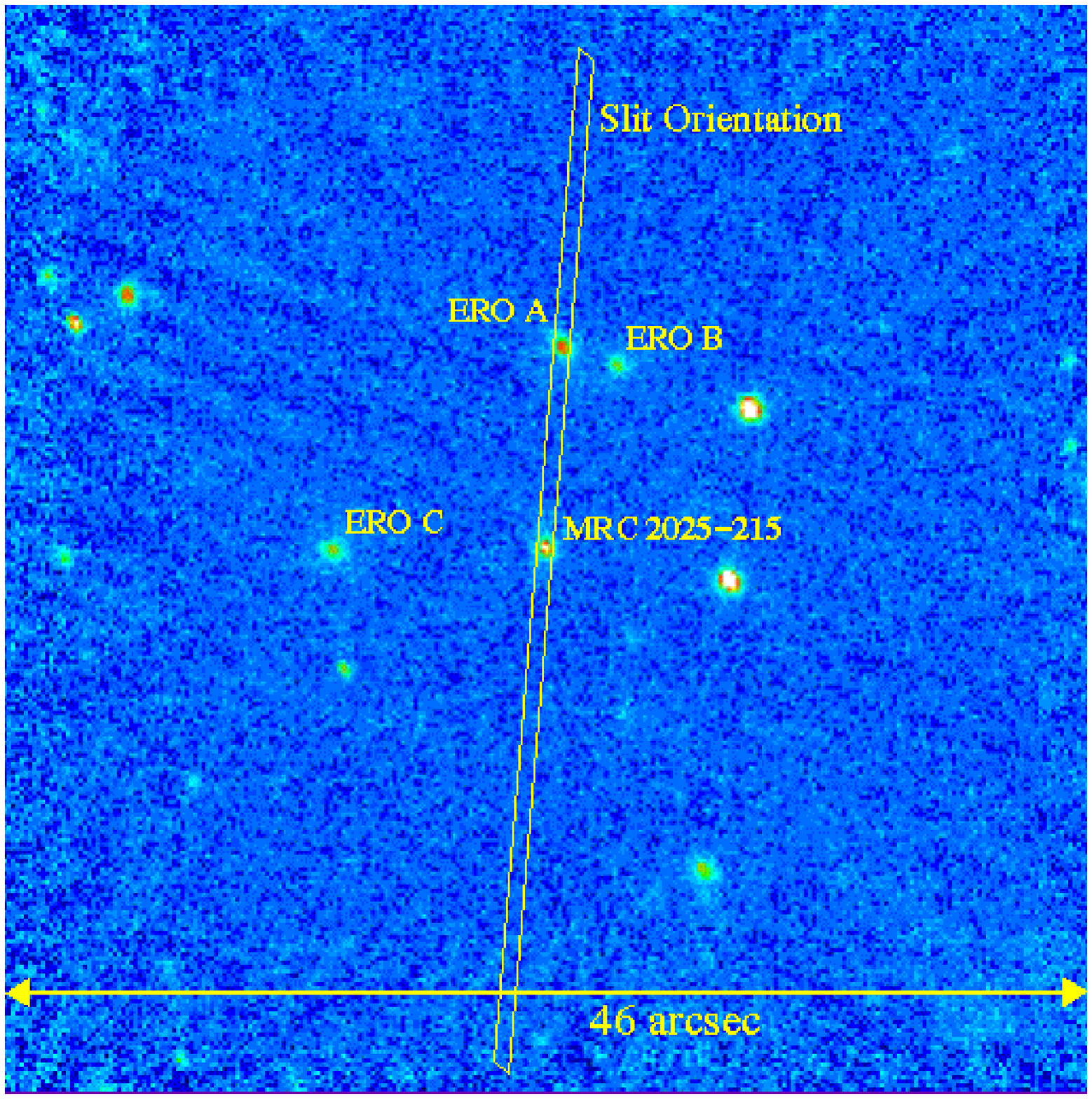}{1.8in}{0}{28}{28}{-195}{-60}
\plotfiddle{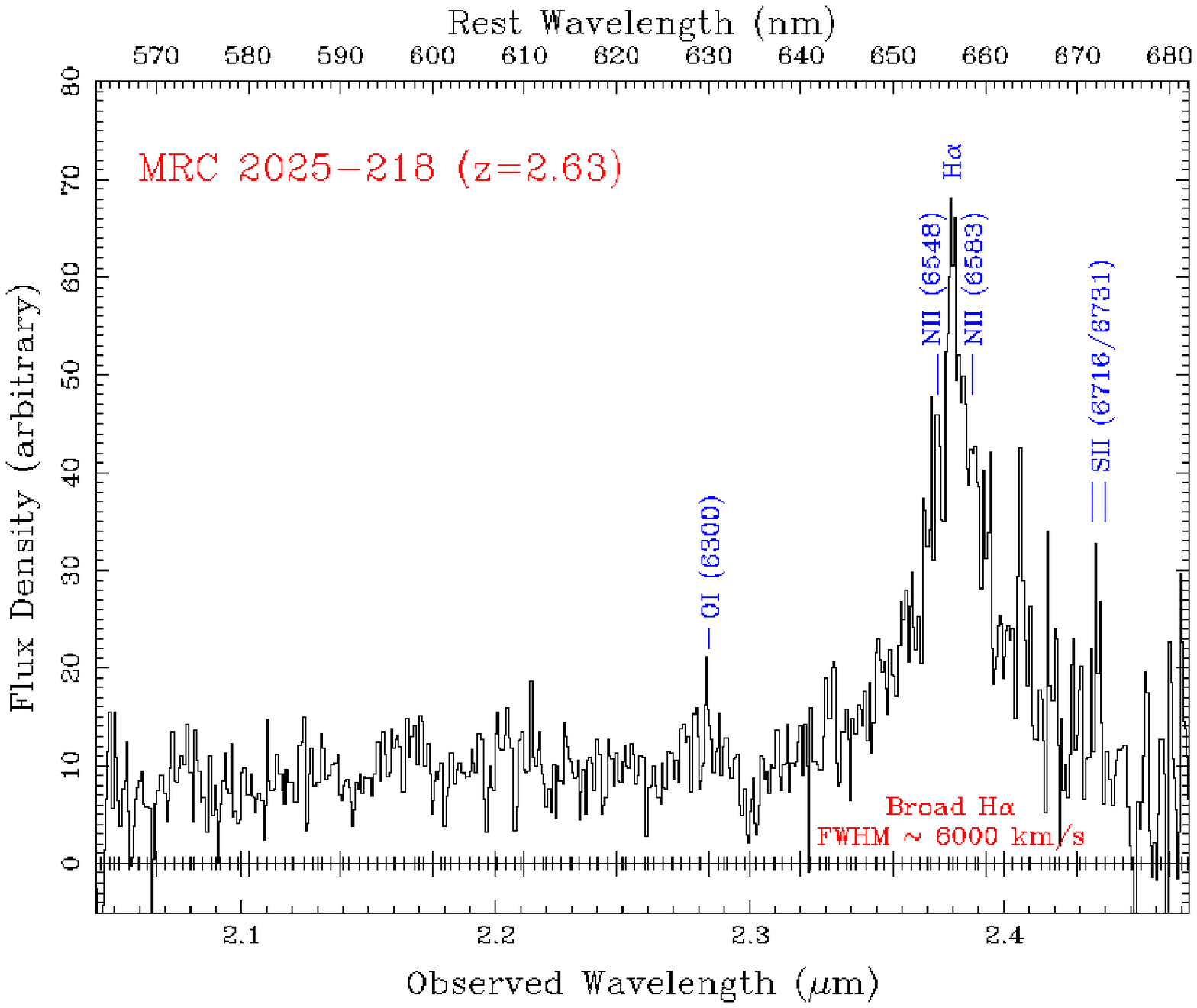}{0.0in}{0}{40}{40}{-35}{-80}
\caption{The K band image of the MRC 2025-218 field is on the left.
The K band spectrum is on the right and is dominated by a very
wide ($>$6000 km s$^{-1}$) strong emission line of H$\alpha$.
\label{fig:image}}
\end{figure}

Four 300 second exposures were taken in both the H-Band
($\sim$1.6$\mu$m) and K-Band ($\sim$2.2$\mu$m).  Due to vignetting at
one edge of the slit, half of the exposures on the ERO were lost so
the effective integration time on MRC 2025-215 is 20 minutes in each
band but only 10 minutes on ERO-A.  The seeing was 0\farcs54.

\section{Results}

Figure 2 shows the H band spectrum of MRC 2025-218.  By far the most
dominant line is [OIII] (500.7 nm) redshifted to 1.82 $\mu$m.  This
line is highlighted in the right panel of figure 2 where the complete
position velocity map of this line is presented. The nucleus has a
double peaked structure in [OIII] and H$\alpha$ (see below) with a
separation of 150 km sec$^{-1}$.  Two knots appear at essential 0 km
sec$^{-1}$ relative velocity, but 1\farcs8 North and 2\farcs4 South of
the Nucleus.  A high speed clump appears 1'' North of the nucleus and
at a redshifted relative velocity of 410 km sec$^{-1}$.  Also detected
in the H band spectrum is the other member of the [OIII] doublet at
495.9 nm, and H$\beta$.  The ratio of [OIII]~/~H$\beta$ is extremely
large at approximately 31$\pm$6.  The H$\beta$ line has a total
nuclear flux of only 3.4~$\times$~10$^{-17}$~ergs~cm$^{-2}$~s$^{-1}$.

\begin{figure}
\plotfiddle{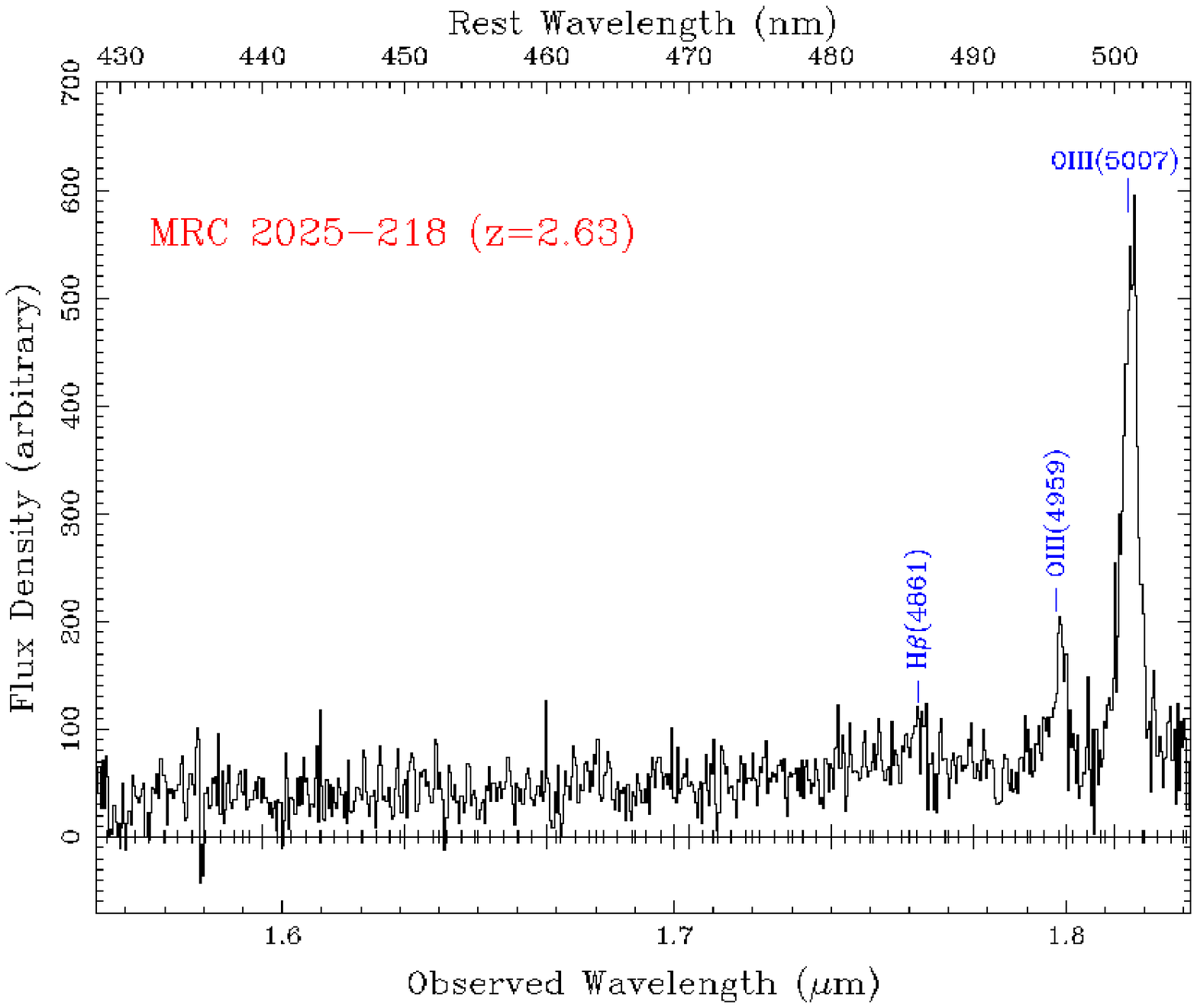}{1.8in}{0}{40}{40}{-210}{-105}
\plotfiddle{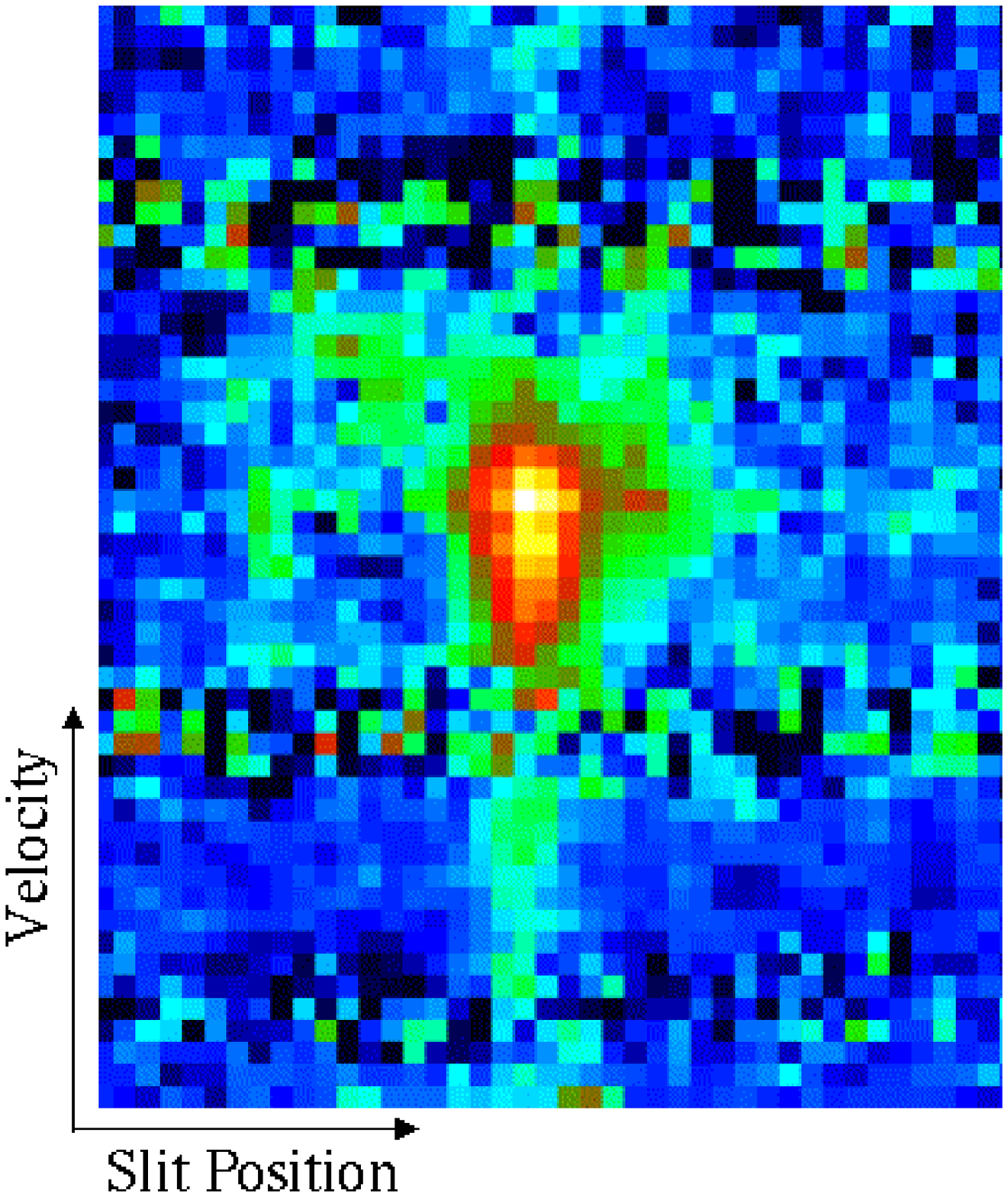}{0.0in}{0}{40}{35}{-20}{-50}
\caption{H band spectrum of MRC 2025-218. The right panel is
the position-velocity plot of just the OIII line at 500.7 nm.
\label{fig:mrch}}
\end{figure}

The right panel of figure 1 shows the K Band spectrum which is
dominated by a broad H$\alpha$ emission line. The line is very
non-Gaussian but has a fwhm of the broad line greater than 6000 km
sec$^{-1}$.  The narrow component has a similar double peaked profile
as the [OIII] line at 500.7 nm. Several other lines are also detected,
but only at the few sigma level, including [OI] (630.0 nm), [NII]
(654.8/658.3 nm) and [SII] (671.6/673.1 nm).  These lines are all much
weaker than H$\alpha$ and the [OI]/H$\alpha$ ratio for the narrow
components is estimated at approximately 0.1 or less.

H and K band spectra of the object labeled ERO-A were taken
simultaneously with MRC 2025-218.  The spectra have no significant
line detections or spectral breaks. In particular no emission lines are
observed at similar positions as MRC 2025-218.  Given the mild H and K
spectral slopes in the spectra and the large r-K color found by
McCarthy et al. (1992), we suggest that the ERO is probably at a
significantly smaller redshift than the radio galaxy.

\section{Discussion \& Conclusions}

The spectrum of the HzRG MRC 2025-218 is clearly dominated by emission lines
from an AGN.  The broad H$\alpha$ emission line is definitive evidence of
a type I AGN.  The extremely large ratio
of [OIII]/H$\beta$ is also found only in true AGN and is in fact greater
than in most local Seyfert 1 type galaxies.

The double peak found in both H$\alpha$ and [OIII] is highly suggestive
of a double active nucleus. If the second peak were due to a star forming
region it would be unlikely that the [OIII] line would be double as well.
The off nucleus knots seen in [OIII] are difficult to understand. Extended
[OIII] has been observed in other radio galaxies aligned to the radio axis
(Armus et al. 1998). Extended Ly$\alpha$ has also been observed aligned
to the radio axis (e.g. Chambers et al., 1996) but the emission mechanism
is poorly understood.  In the case of [OIII] it is difficult to find
a mechanism for its strong production in these off nuclear sites and we
believe that the most likely explanation is that we are seeing [OIII]
as scattered light originally emitted by the nucleus.

The narrow line ratio of H$\alpha$ over H$\beta$ is roughly 8 compared
with the intrinsic ratio of 3.1 observed in local AGN.  If we assume
the difference is due to dust extinction then it is consistent with
obscuration of A$_V$ $\sim$ 2 mag. The fact that H$\beta$ is not
broadened and that broad Ly$\alpha$ was not observed in optical spectra
implies that the central region is much more obscured.  The presence of
such an obscured broad line region in a classic radio galaxy strongly
supports the AGN unification models and links radio galaxies with radio
loud quasars.

The presence of 3 extremely red galaxies discovered within 20'' of the
radio galaxy is very suggestive of a connection with MRC 2025-218. But
if our suggestion is right that the ERO's are at a much lower redshift
then their overdensity in this field and their general overdensity in
the fields of high redshift quasars and radio galaxies may be due to
weak lensing. The connection is that in flux limited surveys you are
more likely to discover high redshift objects if there is an
overdensity of foreground objects.

\acknowledgments

We want to thank the incredibly hard working NIRSPEC instrument team
at UCLA: Maryanne Anglionto, Odvar Bendiksen, George Brims, Leah
Buchholz, John Canfield, Kim Chim, Jonah Hare, Fred Lacayanga, Samuel
B. Larson, Tim Liu, Nick Magnone, Gunnar Skulason, Michael Spencer,
Jason Weiss and Woon Wong.  We would also like to thank the wonderful
scientists and staff members of the Keck Observatory who've made the
commissioning of NIRSPEC extremely productive.  In particular, Tom
Bida, the NIRSPEC support scientist has worked diligently to ensure
the instrument's success.  We also want to thank Lee Armus for useful
discussions.

\end{document}